\documentstyle[12pt]{article}
\nopagebreak

\def\oneh{{\textstyle {1\over 2}}}
\def\oneq{{\textstyle {1\over 4}}}

\def\3j#1#2#3#4#5#6{\mbox{$\left(\begin{array}{ccc}
#1 & #2 & #3 \\
#4 & #5 & #6
\end{array}
\right)$}}

\def\6j#1#2#3#4#5#6{\mbox{$\left\{\begin{array}{ccc}
#1 & #2 & #3 \\
#4 & #5 & #6
\end{array}
\right)$}}

\def\eq{\begin{equation}}

\def\ee{\end{equation}}

\def\eqa{\begin{eqnarray}}

\def\eea{\end{eqnarray}}

\def\ket#1{\mbox{$\vert #1\rangle$}}

\begin{document}


\centerline{\Large{Nuclear Compton scattering in the $\Delta$-resonance
region}}

\medskip

\centerline{\Large{with polarized photons}}

\vskip 1.5cm

\centerline{\large{B.~Pasquini and S.~Boffi}}

\vskip 1.0cm

\centerline{\small Dipartimento di Fisica Nucleare e Teorica, Universit\`a
di Pavia, and}

\centerline{\small Istituto Nazionale di Fisica Nucleare,
Sezione di Pavia, Pavia, Italy}

\vskip 1.5cm


\begin{abstract}

\noindent Nuclear Compton scattering in the $\Delta$-resonance region is
reconsidered within the framework of the $\Delta$-hole model. The different
role of the resonant and non-resonant contributions to the transition
amplitudes is discussed and their effect is investigated by comparing the
results of calculation with recent data also taken with polarized photons.

\end{abstract}

\bigskip




\begin{section}{Introduction}

Photon scattering on nuclei is a pure electromagnetic process providing
a very useful tool for investigating the nuclear structure and dynamics. The
photon probe has a well-known interaction and a mean free path much longer
than the target dimensions, thus exploring the entire target volume. The
scattered photon emerges in general through a two-step mechanism involving
the whole internal dynamics of the target nucleus through virtual
excitations in the intermediate state. In the elastic scattering case, i.e.
Compton scattering, coherence among the different transition amplitudes is
demanded to recover the ground state in the final state. On the one hand,
this implies the possibility of using closure and calculating ground-state
expectation values with a minimum uncertainty since ground-state wave
functions are best known. On the other hand, the role of the transition
operator and the relevant degrees of freedom is emphasized.

Depending on the photon energy, different nuclear degrees of freedom come
into play. At low energy, low-energy theorems can be derived from basic
principles, such as Lorentz invariance and gauge invariance of the
electromagnetic interaction~\cite{[Low],[GG]}. For spin-$\oneh$ particles
the elastic photon scattering amplitude is determined up to first order in
the photon energy by global properties such as its charge, mass and
magnetic moment. In general, up to second order in the photon energy the
nuclear response can be described in terms of the static electric and
magnetic polarizabilities~\cite{[Sachs],[Silbar],[AGreiner]}. A correct
treatment of the c.m. contribution is also important~\cite{[Ericson]} and
must be included consistently~\cite{[Friar]}.

By expanding the elastic photon scattering amplitude into multipole fields
of the incoming as well as the scattered photon, generalized
polarizabilities  can be defined and an extension of the low-energy
theorems has been derived~\cite{[AW]}. A consistent study is then possible
up to and above the pion production threshold~\cite{[WA]} to test the
stressed importance of meson-exchange effects~\cite{[Chris]}.

Cross sections for elastic photon scattering are very small and the energy
resolution in the detection of the scattered photon must be sufficient to
exclude inelastic scattering. Thus, elastic-scattering experiments are
difficult to carry out and existing data are limited and only refer to
zero-spin nuclei. Below pion threshold,  combined analyses of absorption and
scattering  data~\cite{[Bau],[Schela],[Schelb]} were successful in
describing the scattering process in terms of bound-nucleon properties and
in extracting information about the nuclear polarizabilities. Further
references together with a detailed discussion of the theoretical frame and
experimental methods can be found in refs.~\cite{[Vistas]} and
\cite{[Ziegler]}, respectively.

At higher energies, where the $\Delta$ resonance becomes important, early
measurements~\cite{[HZ]} on $^{12}$C and $^{208}$Pb have been followed by an
extensive investigation on $^{4}$He from pion threshold to the $\Delta$
region with bremsstrahlung
beams~\cite{[Austina],[Austinb],[Delli],[Miller]}. In this energy domain
Compton scattering provides information about the intermediate formation and
propagation of the $\Delta$ isobar in the nuclear medium, which is
complementary to that obtained from pion scattering and pion
photoproduction. Different $\Delta$-hole models have been
proposed~\cite{[OsetW],[Weise],[KMO]} to account for a unified description
of such processes. In particular, a consistent picture can be obtained
between data from pion scattering~\cite{[Hirata],[Hiratab],[Lenz]} and pion
photoproduction~\cite{[Kara],[KMa],[KMb]}. The Compton scattering data have
been compared with results obtained within the same $\Delta$-hole
model~\cite{[KMO]}. The data at 320 MeV~\cite{[Austinb]} near the peak of
the $\Delta$ resonance compare well with the predictions of the
$\Delta$-hole model. At lower energies, however, the model fails to
reproduce the magnitude of the cross section at forward angles and the
strong back-angle rise. At backward angles inelastic contributions are
important~\cite{[AWR],[VOTD]} as in the case of pion
photoproduction~\cite{[KO]}. However, as repeatedly noted, also
non-resonant background contributions to coherent scattering are relatively
large outside the $\Delta$-resonance region and may not have been correctly
included in the model. On the other hand, the impulse approximation with
$\gamma$N amplitudes from relativistic dispersion relations, supplemented
by Siegert-like arguments to include the main E1-part of the meson-exchange
contributions~\cite{[Lvov]}, seems to give a reasonable
description of the $^4$He data.

Recently, elastic and inelastic scattering of monochromatic photons from
$^{12}$C was investigated in the energy range across the
$\Delta$ resonance~\cite{[Wissmann]}. The observed elastic cross section at
a scattering angle of 40$^{o}$ was compared with a recent
$\Delta$-hole calculation~\cite{[Osterfeld]} which differs from~\cite{[KMO]}
by a changed pion coupling and by inclusion of the $\rho$-meson into the
final-state interaction. As was also found for $^4$He the $\Delta$-hole
model gives a good agreement in the region of the resonance, but it fails to
reproduce the data below the $\Delta$ peak where they are dominated by a
non-resonant background.

With polarized photons two structure functions contribute to the cross
section~\cite{[Vesper]}. The first structure function, $W_{T}$, is the
same quantity determined by scattering of unpolarized photons and is the
incoherent sum of photon-helicity flip and non-flip contributions. The
second structure function, $W_{TT}$, contains interference contributions
from helicity flip and non-flip amplitudes. Thus the resonant and
non-resonant amplitudes enter quite differently in $W_{T}$ and  $W_{TT}$.
Their separated determination is made possible by a combined
measurement of the differential cross sections for photons linearly
polarized in the reaction plane and perpendicular to it. Such a separation
has been achieved in a recent experiment on $^4$He~\cite{[Schaerf]}
covering the same energy range previously explored with unpolarized
bremsstrahlung beams~\cite{[Austina],[Austinb],[Delli],[Miller]}. This adds
further information to the precise data recently obtained at Mainz with
tagged photons~\cite{[Selke]}.

The enriched quality of the data opens the possibility of a better
understanding of the reaction mechanism. In this paper the problem is
reconsidered within the frame of the $\Delta$-hole model. In order to
reduce the computational effort, the local version of the model developed in
refs.~\cite{[Salcedo],[Garcia],[Nieves]} is adopted to describe the resonant
$\Delta$ formation and propagation. This simplified version retains
essential ingredients of the original $\Delta$-hole model and provides a
useful framework to describe photon absorption~\cite{[CO],[COS]},
pion-nucleus scattering~\cite{[Nieves]}, coherent $\pi^0$ photo- and
electro-production (refs.~\cite{[CNO]} and~\cite{[Hirenzaki]}, respectively)
and Compton scattering~\cite{[Carrasco]}.

In sect. 2, the general formalism for photon scattering is reviewed and the
model used to describe nuclear Compton scattering in the $\Delta$-resonance
region is described in sect. 3. Results for $^{12}$C and $^4$He are
presented and compared with the experimental data in sect. 4. The
conclusions are drawn in the final section.

\end{section}

\begin{section}{General formalism}

The scattering of a photon with momentum ${\vec k}$ and polarization
$\lambda = \pm 1$ into a photon with momentum ${\vec k}'$ and polarization
$\lambda'$ is described by the scattering amplitude
$T^{\lambda'\lambda}(\vec{k}',\vec{k})$. During the scattering
process the nuclear target undergoes a transition from the initial state
$\ket{I_{i} M_{i}}$ to the final state $\ket{I_{f} M_{f}}$.

According to refs.~\cite{[AW],[Vistas]}, a convenient decomposition of
the scattering amplitude is provided in terms of an expansion into
generalized polarizabilities $P^{J}(M^{\nu'}L',M^\nu L;k',k)$. They
correspond to an expansion of the incoming and scattered photon into
multipole fields of order $L$ and $L'$, respectively, with $M^0 = E$ ($\nu =
0$, electric) and $M^1= M$ ($\nu = 1$, magnetic), while the total angular
momentum transferred to the target nucleus is $J$, which is constrained by
the conditions $\vert L-L'\vert\le J\le L+L'$ and $\vert I_{i} - I_{f}
\vert\le J\le I_{i} + I_{f}$. In general, these polarizabilities
are defined by \eqa
& &P_{J}(M^{\nu'}L',M^\nu L;k',k) = \nonumber \\
& & \nonumber \\
& &\quad =  (-)^{L+L'-I_{f}} \,{\hat L}^2\, {\hat {L'}}^2
\sum_{M_{i} M_{f}}\sum_{MM'm}
\oneq \sum_{\lambda\lambda'} \lambda^\nu {\lambda'}^{\,\nu'} \nonumber \\
& & \nonumber \\
& &\qquad \times (-)^{M_{i}}
\3j{I_{f}}{J}{I_{i}}{-M_{f}}{m}{M_{i}}\,
\3j{L}{L'}{J}{M}{M'}{-m} \nonumber \\
& & \nonumber \\
& &\qquad \times \frac{1}{(8\pi^2)^2}
\int{d}R \int{d}R'\,{\cal D}^{L'\ast}_{M'-\lambda'}(R')\,
T^{\lambda'\lambda}_{M_{i}M_{f} }(\vec{k}',\vec{k})\,
{\cal D}^{L\ast}_{M\lambda}(R),
\label{eq:polgen}\\ \nonumber
\eea
where ${\hat L}^2 = 2L+1$, $R$ and $R'$ denote rotations of the quantizations
axis into the direction of $\vec{k}$ and $\vec {k}'$, respectively.

In the case of elastic photon scattering, i.e. Compton scattering, and
for a zero-spin target nucleus ($I_{i}=I_{f}=0$), only the scalar
electric and magnetic polarizabilities, $P_{0}(EL,EL;k,k)$ and
$P_{0}(ML,ML;k,k)$, respectively, survive in eq. (\ref{eq:polgen}).

In the c.m. frame of reference the differential cross section for
scattering of polarized photons on a nucleus reads~\cite{[Vesper]}
\eq
\frac{{d}\sigma}{{d}\Omega} = \frac{k'}{k}\,
\left(\frac {M_{T}}{4\pi E_{T}}\right)^2\,
\frac{1}{2J_{i}+1}
\sum_{\lambda \lambda' \bar{\lambda}}\sum_{M_{i} M_{f}}
(T^{\lambda'\lambda}_{M_{f}M_{i}})\,
\rho_{\lambda\bar{\lambda}}\,
(T^{\lambda'\bar{\lambda}}_{M_{f}M_{i}})^{\ast},
\label{eq:desigmagen}
\ee
where $M_{T}$ is the target mass and $ E_{T} $ is the total c.m.
energy.

The density matrix $\rho_{\lambda\bar{\lambda}}$ describing the
polarization of the incident photon is given by
\eq
\rho_{\lambda\bar{\lambda}}=\frac{1}{2}\left(
\begin{array}{cc}
1 + {\cal P}                        &  - {\cal L} {e}^{-2{i}\phi} \\
- {\cal L} {e}^{2{i}\phi}   &  1 - {\cal P}
\end{array}
\right),
\label{eq:rophoton}
\ee
where $\cal L$ ($\cal P$) is the relative intensity of linearly (circularly)
polarized photons. Assuming the $xz$-plane as the photon scattering plane
with the quantization axis along $\vec k$, $\phi$ is the angle between the
direction of the linear polarization and the $x$-axis.

Using standard Racah algebra, the cross section
(\ref{eq:desigmagen}) becomes
\eqa
\frac{{d}\sigma}{{d}\Omega} &=& \frac{k'}{k}\,
\left(\frac {M_{T}}{4\pi E_{T}}\right)^2\,
\frac{1}{2J_{i}+1}\,
\sum_{\nu \nu'}\sum_{\bar{\nu} \bar{\nu}'}
\sum_{\lambda\lambda'\bar{\lambda}}
\sum_{L L'}\sum_{\bar{L} \bar{L}'}\sum_{J K}\, (-)^{J+K}\,{\hat J}^2\,{\hat
K}^2\, {\cal D}^{K}_{M_K\, 0} \nonumber \\
& &\nonumber \\
& & \times
\lambda^{\nu}{\bar\lambda}^{\bar{\nu}}\lambda'^{\,\nu'+\bar{\nu}'}
\3j{L}{\bar{L}}{K}{-\lambda}{\bar{\lambda}}{-M_K}
\3j{L'}{\bar{L}'}{K}{-\lambda'}{\lambda'}{0}
\6j{L'}{L}{J}{\bar{L}}{\bar{L}'}{K} \nonumber \\
& &\nonumber \\
& &\times
\left[ P_{J}(M^{\bar\nu'}\bar L', M^{\bar\nu} \bar L;k',k) \right]^{\ast}
\rho_{\lambda\bar{\lambda}} \,
P_{J}(M^{\nu'}L',M^\nu L;k',k).
\label{eq:desigpolgen}\\ \nonumber
\eea
Performing the summations over $\lambda$, $\lambda'$ and $\bar{\lambda}$ and
making explicit the $\phi$-de\-pen\-dence of the cross section, one has
\eqa
\frac{{d}\sigma}{{d}\Omega}  & = & \frac{k'}{k}\,
\left(\frac {M_{T}}{4\pi E_{T}}\right)^2\,
\frac{1}{2J_{i}+1}\,
\sum_{\nu \nu'}\sum_{\bar{\nu} \bar{\nu}'}
\sum_{L L'}\sum_{\bar{L} \bar{L}'}\sum_{J K}\, (-)^{J+L+\bar{L}}\,{\hat
J}^2\, {\hat K}^2\,\nonumber \\
& &\nonumber \\
& & \times
\left[ 1 + (-)^{\nu + \bar{\nu} + L + \bar{L} + K} \right]
\3j{L'}{\bar{L}'}{K}{-1}{1}{0}
\6j{L'}{L}{J}{\bar{L}}{\bar{L}'}{K} \nonumber \\
& &\nonumber \\
& & \times
\left[ P_{J}(M^{\bar\nu'}\bar L', M^{\bar\nu} {\bar L};k',k) \right]^{\ast}\,
P_{J}(M^{\nu'}L',M^\nu L;k',k)\nonumber \\
& &\nonumber \\
& & \times\left [
\3j{L}{\bar{L}}{K}{1}{-1}{0} {\cal D}^K_{00} + (-)^{1+\bar{\nu}}
\3j{L}{\bar{L}}{K}{1}{1}{-2} {\cal D}^K_{20}\, {\cal L}\cos 2\phi \right ].
\label{eq:sigpolgen}\\ \nonumber
\eea

Eq. (\ref{eq:sigpolgen}) can also be rewritten as
\eq
\frac{{d}\sigma}{{d}\Omega} = \frac{k'}{k}\,
\left(\frac {M_{T}}{4\pi E_{T}}\right)^2\,
\left( W_{T} + W_{TT} {\cal L} \cos 2\phi\right) ,
\label{eq:cross}
\ee
where the two structure functions are defined in terms of helicity flip
($T^{1-1}_{M_{f}M_{i}}$) and non-flip ($T^{11}_{M_{f}M_{i}}$) amplitudes by
\eqa W_{T} & = &
\frac{1}{2J_{i} +1} \sum_{M_{i} M_{f}}
\left[ \left| T^{11}_{M_{f} M_{i}} \right|^2 +
\left| T^{1-1}_{M_{f} M_{i}} \right|^2 \right] ,
\label{eq:wt}\\
& &\nonumber\\
W_{TT} & = &
-\frac{1}{2J_{i} +1}\sum_{M_{i} M_{f}} 2\,{Re}\,
\left[ T^{11}_{M_{f} M_{i}} (T^{1-1}_{M_{f} M_{i}})^{\ast}
\right].
\label{eq:wtt}\\ \nonumber
\eea

In terms of the generalized polarizabilities (\ref{eq:polgen}), for a
zero-spin target nucleus the Compton scattering amplitude becomes
\eq
T^{\lambda'\lambda}_{M_{i}M_{i}} = \sum_L (-)^{L+1}\hat L^{-1}
{\cal D}^L_{\lambda'\lambda}
\left[ P_0(EL,EL,k,k) + \lambda\lambda'P_0(ML,ML,k,k)\right],
\label{eq:transpol}
\ee
so that the two structure functions are simply given by

\eqa
W_{T} & = &
\sum_{L\bar{L}}(-)^{L+\bar{L}}
{\hat L}^{-1}\,{\hat{\bar{L}}}^{-1}\nonumber\\
& &\nonumber\\
& &
\times\left\{ {\cal D}^{L}_{1,1}(R') {\cal D}^{\bar{L}}_{-1,-1}(R')
\left[ P_0(EL,EL,k,k) + P_0(ML,ML,k,k)\right] \right.\nonumber\\
& &\nonumber\\
& &\qquad \times \left. \left[ P^\ast_0(E\bar{L},E\bar{L},k,k) +
P^\ast_0(M\bar{L},M\bar{L},k,k)\right] \right\}\nonumber\\
& &\nonumber\\
& &\quad
+ {\cal D}^{L}_{1,-1}(R') {\cal D}^{\bar{L}}_{-1,1}(R')
\left[ P_0(EL,EL,k,k) - P_0(ML,ML,k,k)\right] \nonumber\\
& &\nonumber\\
& &\qquad \times \left.
\left[ P^\ast_0(E\bar{L},E\bar{L},k,k) -
P^\ast_0(M\bar{L},M\bar{L},k,k)\right] \right\} ,
\label{eq:wt2}\\
& &\nonumber\\
W_{TT} & = &
- \sum_{L\bar{L}}(-)^{L+\bar{L}}
{\hat L}^{-1}\,{\hat{\bar{L}}^{-1}}\,2\,{Re}\,
\left\{ {\cal D}^{L}_{1,1}(R') {\cal D}^{\bar{L}}_{-1,1}(R')
\right.\nonumber\\
& &\nonumber\\
& & \qquad \times
\left[P_0(EL,EL,k,k) + P_0(ML,ML,k,k)\right] \nonumber\\
& &\nonumber\\
& &\qquad\times\left.
\left[P_0^{\ast}(E\bar{L},E\bar{L},k,k) - P_0^{\ast}(M\bar{L},M\bar{L},k,k)
\right]\right\} .
\label{eq:wtt2}\\ \nonumber
\eea

With linearly polarized photons (${\cal L} = 1$) one can separate these two
structure functions by also measuring the photon asymmetry
\eq
A = \frac{\left({\textstyle{{d}\sigma}  \over
                 \textstyle{{d}\Omega}}\right)_{\parallel} -
          \left({\textstyle{{d}\sigma} \over
                 \textstyle{{d}\Omega}}\right)_{\perp}}
         {\left({\textstyle{{d}\sigma} \over
                 \textstyle{{d}\Omega}}\right)_{\parallel} +
          \left({\textstyle{{d}\sigma} \over
                 \textstyle{{d}\Omega}}\right)_{\perp}}
=\frac{W_{TT}}{W_{T}},
\label{eq:asymm}
\ee
where
\eqa
\left( {\textstyle{{d}\sigma} \over
        \textstyle{{d}\Omega}} \right)_{\parallel}
& = &
\frac{k'}{k}\,
\left(\frac{M_{T}}{4\pi E_{T}}\right)^2 \,
(W_{T} + W_{TT}),
\label{eq:paral}\\
& &\nonumber\\
\left( \frac{\textstyle{{d}\sigma} }
        {\textstyle{{d}\Omega}} \right)_{\perp}
& = &
\frac{k'}{k}\,
\left(\frac{M_{T}}{4\pi E_{T}}\right)^2 \,
(W_{T} - W_{TT})
\label{eq:perpe}\\ \nonumber
\eea
are the cross sections for scattering of photons with polarization
along the $x$-axis ($\phi=0$) and the $y$-axis ($\phi=\pi/2$),
respectively.
\end{section}

\begin{section}{The model}

The electromagnetic interaction Hamiltonian describing photon scattering is
second order in the electromagnetic potential and the scattering amplitude
is the sum of two terms, the one from second-order contributions of the
current-density operator and the other from the first-order contribution
of the genuine two-photon amplitude. In the $\Delta$-resonance region the
first amplitude contains a resonant term with intermediate excitation of a
$\Delta$-hole state which dominates the cross section. All of the other
terms in the total scattering amplitude may be considered as a non-resonant
background. Accordingly, the transition operator $T^{\lambda'\lambda}$ is
split into the sum of a resonant ($R^{\lambda'\lambda'}$) and a background
($B^{\lambda'\lambda}$) part:
\eq
T^{\lambda'\lambda} = R^{\lambda'\lambda} + B^{\lambda'\lambda}.
\label{eq:split}
\ee
In this section a model is described to define $T^{\lambda'\lambda}$ and
explicit expressions for the generalized polarizabilities and the
scattering amplitude are given for the case of a zero-spin nucleus.

\begin{subsection}{The transition operator}

In the original $\Delta$-hole model applied to elastic pion scattering
\cite{[Hirata],[Hiratab],[Lenz]} and coherent $\pi^0$ photoproduction
\cite{[Kara],[KMa],[KMb]}, the $\Delta$ dynamics in the nuclear medium is
dictated by the $\Delta$-hole Green's function
\eq
G_{\Delta {h}}(E) = \left[ E - E_{R}(E) + \oneh {i} \Gamma(E) -
H_{\Delta {h}}\right]^{-1},
\label{eq:Gdeltahole}
\ee
where both the natural free-$\Delta$ width $\Gamma(E)$ and the
resonance energy $E_{R}$ are modified by the effective $\Delta$-nucleus
interaction. Such an interaction is modeled by the Hamiltonian $H_{\Delta
{h}}$ which incorporates $\Delta$-propagation, binding and Pauli
blocking effects as well as a contribution $W_\pi$ describing intermediate
pion propagation in the presence of the nuclear ground state and
corresponding to pion multiple scattering. A complex term, the so-called
spreading potential $V_{sp}$, is also included in $H_{\Delta {h}}$
to account for coupling to multi-hole intermediate channels.

The pion absorption process described by the spreading potential is quite
complicated and a large part of it is expected to be due to the coupling of
the $\Delta$ to the 2p-2h continuum configurations. It has been modeled
either using a phenomenological potential with parameters fitted to the pion
scattering data \cite{[Hirata],[Hiratab],[Lenz]} or performing a microscopic
calculation of the process \cite{[OsetW],[Weise],[Hjort]}.

To reduce the computational effort, a local-density approximation to
the medium-modified $\Delta$ propagator (\ref{eq:Gdeltahole}) has been used
in ref.~\cite{[Kara]} to analyze the pion-nucleus data. The same
approximation has been used in the approach of
refs.~\cite{[Salcedo],[Garcia]} where the $\Delta$-hole Green's function is
written \eq
G_{\Delta {h}}(\rho({\vec r}),{s})
= \left[ \sqrt{s} - M_\Delta +
\oneh {i} \tilde\Gamma(\rho({\vec r}),{s}) -
\Sigma_\Delta(\rho({\vec r}),{s}) \right]^{-1}.
\label{eq:Gdioset}
\ee
In eq. (\ref{eq:Gdioset}) $M_\Delta=1238$ MeV is the $\Delta$ mass, while
$\tilde\Gamma$ and $\Sigma_\Delta$ are the Pauli-blocked $\Delta$ width and
self-energy, respectively. $\Sigma_\Delta$ is in general a non-Hermitean,
non-local and energy-dependent operator. However, when evaluated in the
local-density approximation, both $\tilde\Gamma$ and $\Sigma_\Delta$ become
single-particle operators depending on the nucleon density $\rho({\vec r})$
and the Mandelstam variable $s$ for the photon-nucleon system
\eq
 s = M^2+2\omega\left(M+\frac{3}{5}\,\frac{k_{F}^2}{2M} \right) ,
\label{eq:essebar}
\ee
where $M$ is the nucleon mass, $\omega$ the photon energy and $k_F$ the
(local) Fermi momentum $[k_{F}^{3}={\textstyle{3 \over 2}}\pi ^2\rho
(\vec{r})]$. In eq. (\ref{eq:essebar}) an average over the nucleon momentum
has been performed within the Fermi gas model.
For zero-spin nuclei a spherical nucleon density $\rho(r)$ will be used.

A many-body expansion in terms of ph and $\Delta$h excitations
and the spin-isospin induced interaction has been proposed in
ref.~\cite{[Salcedo]} to evaluate $\Sigma_\Delta$ in nuclear matter
accounting for quasi-elastic corrections, two-body and three-body
absorption. We adopt the same analytical expression and the
corresponding numerical parametrization presented in ref.~\cite{[Salcedo]}
which are supported by the microscopic evaluation of the $\Delta$-spreading
potential in finite nuclei performed in ref.~\cite{[Hjort]}. Moreover, we
include in $\Sigma_\Delta$ the $W_{\pi}$  contribution according to the
indications of ref.~\cite{[KMO]}.

As a consequence of the local-density approximation the part
$R_\Delta^{\lambda'\lambda}$ of the transition operator (\ref{eq:split})
due to the direct excitation of the $\Delta$ resonance becomes
\eq
R_\Delta^{\lambda'\lambda} = \sum_{i=1}^A
F^\dagger_{\gamma{N}\Delta}({\vec k}',i)
G_{\Delta{h}}(\rho({\vec r}_i),{s})
F_{\gamma{N}\Delta}({\vec k},i),
\label{eq:direct}
\ee
where
\eq
F_{\gamma{N}\Delta}({\vec k},i) =
\frac{f_\gamma}{m_\pi}\,
{\vec\epsilon}_\lambda\cdot{\vec k}\times{\vec S}^\dagger\, T_3^\dagger\,
{e}^{{i}{\vec k}\cdot{\vec r}_i}
\label{eq:vertex}
\ee
is the effective $\gamma{N}\Delta$ vertex with
$f_{\gamma}=0.122$~\cite{[CNO]} and $m_\pi$ the pion mass.

We include a second contribution to the resonant part $R^{\lambda'\lambda}$
coming from the crossed $\Delta$-hole excitation. It is given by
\eq
R_{CR}^{\lambda'\lambda} = \sum_{i=1}^A
{F'}^\dagger_{\gamma{N}\Delta}({\vec k},i)
G_{\Delta{h}}(\rho({\vec r}_i),{s}')
F'_{\gamma{N}\Delta}({\vec k}',i),
\label{eq:crossed}
\ee
where
\eq
F'_{\gamma{N}\Delta}({\vec k}',i) =
\frac{f_\gamma}{m_\pi}\,
{\vec\epsilon}^{\,\ast}_{\lambda'}\cdot{\vec k}'\times{\vec S}\,
T_3\, {e}^{-{i}{\vec k}'\cdot{\vec r}_i},
\ee
\label{eq:vertexp}
\eq
{s}' =
M^2-2\omega\left(M+\frac{3}{5}\,
\frac{k_{F}^2}{2M}
\right) .
\label{eq:essebarp}
\ee
Thus
\eq
R^{\lambda'\lambda} = R_\Delta^{\lambda'\lambda} +
R_{CR}^{\lambda'\lambda} .
\label{eq:resonant}
\ee

Non-resonant background contributions are due to s-wave pion production
and absorption on a single nucleon~\cite{[KMO]} and to a two-photon contact
interaction described by a seagull diagram arising from the quadratic term
in the vector potential of the non-relativistic interaction
Hamiltonian~\cite{[Aren],[WA]}.

Assuming the Kroll-Ruderman form for pion production as in
ref.~\cite{[KMO]}, the corresponding transition operator is
\eq
B_{KR}^{\lambda'\lambda} = \sum_{i=1}^A B(E) \,
{\vec\epsilon}^{\,\ast}_{\lambda'}\cdot{\vec\sigma}(i)\,
{\vec\epsilon}_\lambda\cdot{\vec\sigma}(i)\,
{e}^{{i}({\vec k}-{\vec k}')\cdot{\vec r}_i},
\label{eq:swave}
\ee
where $B(E)$ has been derived by a dispersion integral at the energy
$E=\sqrt{s}$. Equivalently, $E = \omega_{q} + \epsilon_{q}=
\sqrt{m_\pi^2+q^2} + \sqrt{M^2_{} + q^2}$, with $q$ being the c.m. on-shell
momentum in the $\pi$N channel. One has \eq
B(E) = 4\pi\alpha\left(\frac{f}{m_\pi}\right)^2
\int\frac{ {d}{\vec q}{\,}' }{(2\pi)^3}
\frac{M}{\omega_{q'}\epsilon_{q'}}
\left | \frac{v(q'^2)}{v(q^2)}\right | ^2
\frac{1}{E - (\omega_{q'} +
\epsilon_{q'}) + {i}\epsilon},  \label{eq:bidie}
\ee
\eq
{Im\,}B(E) =
- 2\alpha\left(\frac{f}{m_\pi}\right)^2\left(\frac{M}{E}\right) q,
\label{eq:bidiedue}
\ee
with $f^2/4\pi = 0.08$ and $\alpha=e^2/4\pi=1/137$.
The form factor $v(q^2)$ governing the off-shell behaviour is parametrized
as in ref.~\cite{[KMO]}, with $v(q^2)=(1+\beta^2/q^2)^{-1}$ and
$\beta=300$ MeV.
The two-photon operator includes a term describing the Thomson scattering by
individual nucleons and a term due to two-body exchange
currents~\cite{[Aren]}. Both contributions are required to fulfil gauge
invariance. However, in ref.~\cite{[WA]} the exchange contribution was shown
to be rather small. Therefore it will be neglected here and the two-photon
operator is simply given by
\eq B_{TP}^{\lambda'\lambda} =
\frac{1}{M} {\vec\epsilon}^{\,\ast}_{\lambda'}\cdot
{\vec\epsilon}_\lambda
\sum_{i=1}^A  e_i^2\,{e}^{{i}({\vec k}-{\vec k}')\cdot{\vec r}_i},
\label{eq:tpo}
\ee
where $e_i=\oneh[1+\tau_3(i)]e$ is the nucleon electric charge.

Second order contributions from the one-body current with a nucleon in the
intermediate state are rather small in the energy range considered here
and will also be neglected. Therefore the total background part of the
transition operator is
\eq
B^{\lambda'\lambda} =
B_{KR}^{\lambda'\lambda} + B_{TP}^{\lambda'\lambda}.
\label{eq:background}
\ee

Within the same local-density approximation to the $\Delta$ propagator, eq.
(\ref{eq:Gdioset}), the original $\Delta$-hole model of ref.~\cite{[KMO]} is
recovered with the following modifications in the transition operator.

\noindent $i$) In the effective $\gamma{N}\Delta$ vertex of eq.
(\ref{eq:vertex}) the coupling $f_\gamma/m_\pi$ is replaced by
$g_{\gamma{N}\Delta}/M_\Delta$ and an additional factor $V$ is
introduced, which includes a background contribution describing the pion
photoproduction followed by  the resonant pion rescattering (Fig. 1). We
have
\eq
{V}=1-a_B(E) \Sigma^{\pi{N}}_{\Delta},
\qquad
a_B(E)=\frac{\tilde{g}_{\gamma{N}\Delta}}
{g_{\gamma{N}\Delta}}\,
\frac{\sin \phi(E)}{\Gamma(E)/2},
\ee
where $\Sigma^{\pi{N}}_{\Delta}$ is the medium-corrected self-energy
corresponding to intermediate coupling to the $\pi {N}$  channel,
$\tilde{g}_{\gamma{N}\Delta}=1.03$, $g_{\gamma{N}\Delta}=1.02$ and
$\phi(E)$ is parametrized according to ref.~\cite{[KMO]}.

\noindent $ii$) The crossed $\Delta$-hole excitation of eq.
(\ref{eq:crossed}) is replaced by a background photopion production in the
resonant channel
\eq
R_B^{\lambda'\lambda} = \sum_{i=1}^A
F^\dagger_{\gamma{N}\Delta}({\vec k}',i)
A_B(E)
F_{\gamma{N}\Delta}({\vec k},i),
\ee
where
\eq
A_B(E)=a_B^2(E)\Sigma_{\Delta}^{\pi {N}}.
\label{eq:abdie}
\ee

\noindent $iii$) The non-resonant background is simply given by the
transition operator  $B_{KR}^{\lambda'\lambda}$ of eq. (\ref{eq:swave})
with  $B(E)\to B'(E)$, which is obtained by introducing under the
integral of eq. (\ref{eq:bidie}) a phenomenological energy-dependent
factor $h^2(E)$ fitted to the total  photo-absorption cross section.

\end{subsection}

\begin{subsection}{Generalized polarizabilities for zero-spin nuclei}

The transition operator for photon scattering in the model described above
turns out to be a single-particle operator. When evaluating the
corresponding transition amplitude all matrix elements can be expressed
in terms of the nuclear-matter density. Therefore the nuclear density enter
twice, first in the local-density approximation to the $\Delta$-hole
propagator and second as a result of an effective impulse approximation.

Confining the discussion to zero-spin nuclei, only the scalar electric and
magnetic polarizabilities are different from zero. The resonant contribution
to the polarizabilities is given by

\eqa
& & P^{{R}}_0(EL,EL;k,k) =
\left(\frac{f_\gamma}{m_\pi}\right)^2
\frac{2}{9}\, \tilde{k}\tilde{k}'\,(-)^{L+1}\hat{L} \nonumber\\
& & \qquad \times \int{d}{\vec r}\,
\rho(r) (2L+1)\, j^2_{L}(kr)\,\left[G_{\Delta{h}}(\rho(r),{s})+
G_{\Delta{h}}(\rho(r),{s'})\right] ,\nonumber\\
& & \\
& & \nonumber\\
& & P^{{R}}_0(ML,ML;k,k) =
\left(\frac{f_\gamma}{m_\pi}\right)^2
\frac{2}{9}\, \tilde{k}\tilde{k}'\,(-)^{L+1}\hat{L} \nonumber\\
& &\quad\times\int{d}{\vec r}\, \rho(r)\,
\left[ L\, j^2_{L+1}(kr) + (L+1)\,j^2_{L-1}(kr)\right] \nonumber\\
& &\qquad\times[G_{\Delta{h}}(\rho(r),{s}) +
G_{\Delta{h}}(\rho(r),{s'})],
\label{eq:pdeltamag} \\ \nonumber
\eea
where $\tilde{k}$ is the photon energy in the photon-nucleon
system~\cite{[CNO]}.

The other background contributions are
\eqa
& &P^{{B}}_0(EL,EL;k,k) =
\left[\oneh B(E) + \frac{e^2}{4 M}\right]
 (-)^{L+1}\hat{L} \nonumber\\
& &\qquad \times\int{d}{\vec r}\, \rho(r) \,
\left[ L\, j^2_{L+1}(kr) + (L+1)\,j^2_{L-1}(kr)\right] ,
\label{eq:pstpoel} \\
& & \nonumber\\
& &P^{{B}}_0(ML,ML;k,k) =
\left[\oneh B(E) + \frac{e^2}{4 M}\right]
(-)^{L+1}\hat{L}  \nonumber\\
& &\qquad\qquad \times
\int{d}{\vec r} \,\rho(r)\, (2 L+1)\, j^2_{L+1}(kr) .
\label{eq:pstpomag} \\ \nonumber
\eea

The corresponding expressions in the local-density approximation to the
original $\Delta$-hole model of ref.~\cite{[KMO]} are obtained with the
modifications $i$)--$iii$) explained in sect. 3.1.

\end{subsection}

\begin{subsection}{Compton scattering amplitude}

The Compton scattering amplitude for zero-spin nuclei in the c.m. frame
becomes
\eqa
T^{\lambda'\lambda}
& = & \frac{1 + \lambda\lambda'\cos\theta}{2}
\left\{ F(q^2)\left[ B(E) + \frac{e^2}{2 M} \right] \right. \nonumber \\
& & \label{eq:totaltrans} \\ \nonumber
& &{} + \left. \frac{4}{9}\left(\frac{f_\gamma}{m_\pi}\right)^2
\tilde{k}\tilde{k}'
\int{d}{\vec r} \, {e}^{{i}({\vec k} - {\vec k}')\cdot{\vec r}}
\rho(r)
\left[ G_{\Delta{h}}(\rho(r),{s}) +
G_{\Delta{h}}(\rho(r),{s}')  \right]
\right\} ,
\\ \nonumber
\eea
where
\eq
F(q^2) = \int{d}{\vec r} \,
{e}^{{i}({\vec k} - {\vec k}')\cdot{\vec r}} \rho(r)
\label{eq:ff}
\ee
is the nuclear form factor with ${\vec q} = \vec{k} - \vec{k}'$.

In the local-density approximation to the original $\Delta$-hole model of
ref.~\cite{[KMO]} discussed in sect. 3.1 the Compton scattering amplitude
(\ref{eq:totaltrans}) becomes
\eqa
T^{\lambda'\lambda}
& = & \frac{1 + \lambda\lambda'\cos\theta}{2}
\left\{ F(q^2) B'(E) +
\frac{4}{9} \tilde{k}\tilde{k}'
\int{d}{\vec r} \, {e}^{{i}({\vec k} - {\vec k}')\cdot{\vec r}}
\rho(r)
 \right. \nonumber \\
& & \label{eq:totaltranskmo} \\ \nonumber
& &\qquad\times \left. \left[
\left(\frac{g_{\gamma{N}\Delta}}{M_\Delta}V\right)^2 \,
G_{\Delta{h}}(\rho(r),{s})+
\left(\frac{g_{\gamma{N}\Delta}}{M_\Delta}\right)^2 \,
A_{B}(E) \right]
\right\} .
\\ \nonumber
\eea

\end{subsection}

\end{section}

\begin{section}{Results}

The model is applied to study Compton scattering from zero-spin
nuclei such as $^4$He and $^{12}$C. The nuclear-matter density for such
nuclei can be identified with the charge density derived from the charge
form factor measured  with high accuracy in elastic electron scattering.

Recently, new data on $^{12}$C have been accumulated at a scattering angle
$\theta=40^{o}$ in the photon energy range between 200 and 500
MeV~\cite{[Wissmann]}. At the highest energy the corresponding momentum
transferred to the recoiling target nucleus reaches the value of the first
minimum in the charge form factor of $^{12}$C ($q\approx 1.7$ fm$^{-1}$).
This kinematical situation makes it possible to test the validity of the
local-density approximation to the $\Delta$-hole model.

In fig. 2 the data are shown together with different model calculations.
The dotted curve is the result of a full $\Delta$-hole
calculation~\cite{[Osterfeld]}, where the residual $\Delta$-hole
interaction includes $\pi$ and $\rho$ exchanges as well as short-range
correlations simulated by the Landau-Migdal parameter. Including the
residual interaction, a good description of the data is obtained at photon
energies higher than 250 MeV. The discrepancy between theory and data at
lower energies has been ascribed to the importance of background
effects~\cite{[Osterfeld]}.

The solid curve is obtained with the scattering amplitude
(\ref{eq:totaltranskmo}) corresponding to the original $\Delta$-hole
model~\cite{[KMO]} treated in the local-density approximation. The adopted
nuclear density is derived within a projected-Hartree-Fock approach to the
description of the ground state properties of $^{12}$C and accurately
accounts for the charge form factor all over the explored range of
momenta~\cite{[Bouten]}. The difference between the solid and dotted curves
is a combined effect of the local-density approximation and the medium
modifications applied to the effective $\gamma{N}\Delta$ vertex and
to the background contributions according to ref.~\cite{[KMO]} and not
included in the calculation of ref.~\cite{[Osterfeld]}.

The dashed curve corresponds to the scattering amplitude
(\ref{eq:totaltrans}) with the same nuclear density used for the solid
curve. The different background and the contribution of the crossed
$\Delta$-hole excitation reduce the size of the peak and shift its position
to higher energies.

According to the results of fig. 2 the uncertainty introduced by the
local-density approximation in the case of Compton scattering is of the
same magnitude as that obtained in coherent pion production~\cite{[CNO]}.
The main difference between the model calculations stems from the different
ingredients taken into account in the transition mechanism. In particular,
comparison of the different results with data in fig. 2 confirms the
important role of the background at the lower energies. A detailed study of
it can be done with the aid of the recent data on $^4$He obtained at
Mainz~\cite{[Selke]}. In the explored photon energy range between 200 and
500 MeV at fixed scattering angle ($\theta_{cm}=40^{o}$) the
momentum transferred to the recoiling target nucleus corresponds to the
low-$q$ part of the charge form factor where a gaussian-shaped $\rho(r)$ is
a good approximation to the nuclear density. Furthermore, the relatively
small variation of $F(q^2)$ in this range gives a better insight into the
role of the other theoretical ingredients.

In figs. 3 and 4 the results are shown as obtained in the local-density
approximation to the $\Delta$-hole of ref.~\cite{[KMO]} and with the
scattering amplitude (\ref{eq:totaltrans}), respectively. As expected, the
background contribution is smooth in both cases as a consequence of the
small variation of $F(q^2)$ with $q$ and the weak energy-dependence of the
non-resonant amplitudes $B(E)$ and $B'(E)$. However, a much larger
background is produced by including the two-photon operator in the
scattering amplitude (\ref{eq:totaltrans}) as required by gauge invariance.
Its effect is hardly taken into account by the modification
$B(E)\to B'(E)$ introduced in ref.~\cite{[KMO]} whose limitations were
already stressed by the authors themselves. On the other hand, a quite
different resonant contribution is provided by the two models. Here the
medium effects are substantial in defining the correct position of the peak
energy. While these effects were accurately taken into account in
ref.~\cite{[KMO]}, they are only in part considered in the calculation
with the scattering amplitude (\ref{eq:totaltrans}). In addition, the
crossed $\Delta$-hole excitation shifts to higher energies the peak
location, also anomalously increasing the resonant-background interference.
In any case, up to about 350 MeV both models account for the data
satisfactorily, thus giving confidence that a local-density approximation
is well suited to by-pass the technical difficulties connected with the
full $\Delta$-hole calculation.

For the first time Compton scattering data with polarized photons have
become available~\cite{[Schaerf]}. The experiment was performed on $^4$He
by the LEGS collaboration at Brookhaven in the energy range between 200
and 310 MeV at five different incoming energies. The data for three of
them are shown in fig. 5 as a function of the c.m. scattering angle for a
photon beam fully polarized in the reaction plane ($\phi=0^{o}$,
$({d}\sigma/{d}\Omega)_{\parallel}$ in eq.~(\ref{eq:paral})) and
perpendicular to the reaction plane ($\phi=90^{o}$, $({d}
\sigma/{d}\Omega)_{\perp}$ in eq. (\ref{eq:perpe})). Quite
similar results are obtained in the local-density approximation to the
$\Delta$-hole of ref.~\cite{[KMO]} (solid curve) and with the scattering
amplitude (\ref{eq:totaltrans}) (dashed curve). The observed behaviour at
forward angles is well reproduced, while discrepancies are present in the
backward scattering region. This is a well known problem already stressed
in ref.~\cite{[Miller]} and sometimes related to an unsatisfactory
treatment of background contributions~\cite{[Delli]}.

With polarized photons it is possible to separate the two structure
functions $W_{T}$ and $W_{TT}$ which are differently related to
the resonant and background contributions. This separation is illustrated
in fig. 6, where the unpolarized ($W_0$) and polarized ($W_1$) cross
sections for elastic Compton scattering off $^4$He are reported. They are
given in terms of the two structure functions $W_{T}$ and $W_{TT}$
as follows:
\eq
W_0 = \frac{k'}{k}\,
\left(\frac {M_{T}}{4\pi E_{T}}\right)^2\,W_{T}, \qquad
W_1 = \frac{k'}{k}\,
\left(\frac {M_{T}}{4\pi E_{T}}\right)^2\,W_{TT} .
\label{eq:ww}
\ee
In turn, taking benefit of the local-density approximation the structure
of $W_{T}$ and $W_{TT}$ is quite simple:
\eq
W_{T} = \frac{1}{2} (1 + \cos^2\theta_{cm})
\left[ \vert r(E)\vert^2 + \vert b(E)\vert^2 \right]
+ 2\cos\theta_{cm}\, {Re}\, \left[ r(E) b^*(E)\right] ,
\label{eq:wtp}
\ee
\eq
W_{TT} = \frac{1}{2}\sin^2\theta_{cm}
\left[ \vert r(E)\vert^2 - \vert b(E)\vert^2 \right],
\label{eq:wttp}
\ee
where $r(E)$ and $b(E)$ represent the model resonant and background
amplitudes, respectively. Therefore, according to eqs. (\ref{eq:paral})
and (\ref{eq:perpe}), at $\theta_{cm}=90^{o}$,  $({d}\sigma/{d}
\Omega)_{\parallel}\propto \vert r(E)\vert^2$ and $({d}\sigma/
{d}\Omega)_{\perp} \propto \vert b(E)\vert^2$, so that the two cross
sections are uniquely determined by either the resonant or background
contributions. The magnitude of these two terms appears in agreement with
data in fig. 5. On the other hand, the angular behaviour of $({d}
\sigma/{d}\Omega)_{\parallel}$ is determined by the background
through terms in $\cos^2\theta_{cm}$ (pure background) and
$\cos\theta_{cm}$ (background-resonant interference). At the resonance
energy a fair agreement is obtained, while the deficiencies at backward
angles for energies below the $\Delta$-resonance region  show that some
mechanism, whose effects increase with $q$, is lacking there. This is
confirmed by looking at $W_0$ in fig. 6, where the forward-backward
asymmetry is entirely due to the interference term proportional to
$\cos\theta_{cm}$. In the case of $({d}\sigma/{d}
\Omega)_{\perp}$ the role of background and resonant contributions is
interchanged. In the forward emisphere the correct behaviour of $({d}
\sigma/{d}\Omega)_{\perp}$ is mainly determined by the form factor
$F(q^2)$, but again at backward angles the yield is too low. The role of
$F(q^2)$ is better seen in $W_1$ in fig. 6. The $\sin^2\theta_{cm}$
dependence of $W_1$ is modulated by $F(q^2)$ with the result that the peak
in the angular distribution of $W_1$ is shifted at angles lower than
$90^{o}$. In addition, the magnitude of $W_1$ is larger when the
background contribution is smaller.

The behaviour of the two structure functions $W_{T}$ and $W_{TT}$
also determines the photon asymmetry. The peak in the angular distribution
at the different photon energies shown in fig. 7 always occurs at angles
larger than those of the data as a result of the too low response $W_{T}$
at backward angles.

\end{section}

\begin{section}{Conclusions}

The high quality of recent data on elastic Compton scattering off
zero-spin nuclei such as $^4$He and $^{12}$C in the $\Delta$-resonance
region and the possibility of disentangling different nuclear responses
with polarized photons have made possible a detailed investigation of the
interplay between resonant and background contributions in the
interpretation of data within the $\Delta$-hole model. A local-density
approximation has been shown to give reasonable results also for such a
light nucleus as $^4$He, thus giving confidence to be of help when dealing
with heavier targets. Under resonance conditions a rather satisfactory
agreement between theory and data is gained. Below the $\Delta$-resonance
energy discrepancies persist at backward angles which can be ascribed
to some lacking mechanism active in the structure function $W_{T}$.

\end{section}


\bigskip

We are grateful to Carlo Schaerf and Annalisa D'Angelo for useful
discussions and to them as well as to Martin Schumacher and Reinhard Beck
for making available to us their preliminar data.

\clearpage


\clearpage


\centerline{Figure captions}

\medskip

Fig. 1. Effective $\gamma{N}\Delta$ vertex in the background production.

\smallskip

Fig. 2. Differential cross sections for elastic Compton scattering off
$^{12}$C as a function of the incoming photon energy $\omega$ at a fixed
photon angle $\theta=40^{o}$. The data are taken from
ref.~\cite{[Wissmann]}. The dotted curve represents the results of a full
$\Delta$-hole calculation~\cite{[Osterfeld]}. The solid curve is the
local-density approximation to the $\Delta$-hole of ref.~\cite{[KMO]}. The
dashed curve is obtained with the scattering amplitude
(\ref{eq:totaltrans}).

\smallskip

Fig. 3. Differential cross sections for elastic Compton scattering off
$^4$He as a function of the incoming-photon energy $\omega$ at a fixed
c.m. photon angle $\theta_{cm}=40^{o}$. The data are taken from
ref.~\cite{[Selke]}. The solid curve is the local-density approximation to
the $\Delta$-hole of ref.~\cite{[KMO]}. The dashed, dot-dashed and dotted
lines give the separate contributions of the resonant, background and
interference parts, respectively.

\smallskip

Fig. 4. The same as in fig. 3 but for the model described by the
scattering amplitude (\ref{eq:totaltrans}).

\smallskip

Fig. 5. Differential cross sections for elastic Compton scattering off
$^4$He as a function of the c.m. scattering angle for a photon beam fully
polarized in the reaction plane ($\phi=0^{o}$) and perpendicular to the
reaction plane ($\phi=90^{o}$) at the indicated values of the
laboratory photon energy. The data are taken from~\cite{[Schaerf]}. The
solid curve is the local-density approximation to the $\Delta$-hole of
ref.~\cite{[KMO]}. The dashed curve is obtained with the model described
by the scattering amplitude (\ref{eq:totaltrans}).

\smallskip

Fig. 6. The unpolarized ($W_0$) and polarized ($W_1$) cross sections for
elastic Compton scattering off $^4$He as a function of the c.m. scattering
angle at the indicated values of the laboratory photon energy. The data are
taken from~\cite{[Schaerf]}. Solid and dashed curves as in fig. 5.

\smallskip

Fig. 7. The photon asymmetry for elastic Compton scattering off $^4$He as a
function of the c.m. scattering angle at the indicated values of the
laboratory photon energy. The data are taken from~\cite{[Schaerf]}. Solid
and dashed curves as in fig. 5.


\end{document}